\documentstyle[11pt,epic]{article}

\def\beq{\begin{equation}}
\def\eeq{\end{equation}}
\def\bea{\begin{eqnarray}}
\def\eea{\end{eqnarray}}

\def\nn{\nonumber}

\title
{\LARGE\bf The Vertex Corrections  \\ in Technicolor Model \\
without\\  Exact Custodial Symmetry }
\bigskip
\bigskip
\vspace{1cm}
\author{Tadashi YOSHIKAWA  \\
                             \\
\normalsize \em Department of Physics, Hiroshima University\\
\normalsize \em 1-3-1 Kagamiyama, Higashi-Hiroshima, 724\\
\normalsize \em Japan}

\date{}
\begin{document}
\setlength{\baselineskip}{24pt}
\maketitle
\begin{picture}(0,0)
\put(325,240){HUPD-9415}
\put(325,225){November, 1994}
\end{picture}
\vspace{-24pt}

\vspace{3cm}
\begin{abstract} \large
We discuss the effects of isospin breaking which appear in the vertex
corrections for $Zb \bar{b}$, $Z\tau^+\tau^-$ and
$W\nu\tau$ in a one-family technicolor model without exact
custodial symmetry. By means of the effective lagrangian approach
we compute the vertex corrections for $Zb\bar{b}$,
$Z\tau\tau$ and $W\tau\nu$ taking account of
the contributions from technivectormesons. If the isospin
symmetry in technilepton sector is not exact,
technivectormesons contribute to the vertex correction
for $Z\tau\tau$ but such contributions to the correction for
$W\tau\nu$ are absent. If the difference is measured,
it is the evidence of the isospin breaking.
\end{abstract}
\newpage
%%%%%%%%%%%%%%%%%%%%%%%%%%%%%%%%%%%%%%%%%%%%%%%%%%%%%%%%%%%
%                  Main pert
%%%%%%%%%%%%%%%%%%%%%%%%%%%%%%%%%%%%%%%%%%%%%%%%%%%%%%%%%%%%
In the previous work \cite{yosi}, we constructed the
effective Lagrangian for a one-family technicolor model
without exact custodial symmetry and discussed the constraints for
the oblique corrections. The most distinctive feature from
the traditional technicolor theory is the isospin breaking in
technilepton sector\cite{apel}. In the model
\cite{apel}\cite{yosi}, we find that
the constraints
for oblique corrections can be satisfied. The effects of the isospin
breaking must appear in the vertex correction (non-oblique
correction), too.
In this
letter, we
study the vertex correction for $Zbb$, $Z\tau\tau$ and
$W\tau\nu$ in the technicolor model including
isospin breaking with the effective lagrangian.
The corrections depend
on the decay constant of technipion in each sector
( techniquark sector, technilepton sector ) \cite{civ}\cite{kit}\cite{kit2}.
In the model \cite{apel}\cite{yosi}, one of the isospin
breaking effects appears in the
difference between the decay constant of the charged
technipion
and that of neutral technipion in the technilepton sector. From
the constraint of the oblique correction $T$, the difference
between the decay constants of the technipion in technilepton
sector must be enough small compared with decay constant in
the techniquark sector. Their differences
directly appear in the differences between the vertex corrections.
The other larger effect of the isospin
breaking in the technilepton sector comes from the
technivectormesons composed of technileptons.
In the model, there are the neutral technivectormesons which
contribute to the vertex correction for the $Z\tau\tau$,
while the charged technivectormesons which should contribute
to the correction for the $W\tau\nu$ are absent.
Hence,
we may gain some hint about the evidence of isospin
breaking in technilepton sector
through the difference of the vertex corrections between
$Zbb$ and $Z\tau\tau$ as well as the difference
between $Z\tau\tau$ and $W\tau\nu$
in the precision measurements.

%In \cite{civ}, \cite{kit} and \cite{kit2}, the authors
%used a technic which
%replaces the current
%of technifermion by a Noether (chiral) current associated
%with the $SU(2)_L$ symmetry in an effective sigma model \cite{gio}
%\bea
% \bar{T}_L \gamma_\mu \tau^A T_L \rightarrow \frac{F^2}{4} tr
%[ \Sigma^\dagger \tau^A D_\mu \Sigma ]
%\eea
%where $\Sigma = exp(2i\pi^a \tau^a / F )$ , $\pi^a$ show pions
%and F is a pion decay constant.
%So we also use same technic to compute it, that is, we replace
%the current in four fermi interaction to Noether current in our effective
%Laglangian \cite{yosi}.

The vertex corrections depend on the Extended Technicolor
Model ( ETC ). The Lagrangian \footnote{Similarly some
diagonal
ETC gauge interactions between the
same family also exist \cite{kit2}.
The vertex
corrections for this interaction as depicted in Fig.1(b) is
exist.
The vertex is
effectively same with Fig.2, except for the order of technicolor's number
$N_{tc}$. By mean of $1/N$ expansion we can ignore this
dependence, but if $N_{TC}$ is small, we need to consider
the effects. }
which describes the ETC gauge
interaction of one family technicolor model
between  the third family and Technifermion
is,
\bea
   {{\cal L}_{ETC (3-TC)}}
                       &=&  g_{ETC} \xi_L^t \bar{Q}_L^i
                            W_{ETC}^\mu \gamma_\mu q_L^i  \nn \\
                         &+& g_{ETC} \xi_R^t \bar{t}_R^i
                            W_{ETC}^\mu \gamma_\mu U_R^i
                        + g_{ETC} \xi_R^b \bar{b}_R^i
                            W_{ETC}^\mu \gamma_\mu D_R^i + h.c. \nn \\
                        &+&  g_{ETC} \xi_L^\tau \bar{L}_L
                           W_{ETC}^\mu \gamma_\mu l_L  \nn \\
                        &+& g_{ETC} \xi_R^\tau \bar{\tau}_R
                              W_{ETC}^\mu \gamma_\mu E_R
                              + h.c. ,
\label{3-tc}
\eea
where $Q_L^i = ( U^i,D^i )_L$, $U_R^i$ and $D_R^i$ represent
techniquarks,
$q_L^i = ( t^i,b^i )_L$, $t_R^i$ and $b_R^i$ represent
the third family of quarks and `` i ''
is the color index of QCD. $L_L = ( N,E )_L$, $E_R$
represent the technilepton, $l_L = ( \nu,\tau )_L$
and $\tau_R$ represent the third family of leptons. $g_{ETC}$
is a coupling of ETC interaction. $W_{ETC}$
is an ETC gauge boson which mediates between the third
family of ordinary fermions and techni
fermions.
$\xi_L^{t(\tau)}$ is a coefficient of left handed
coupling and $\xi_R^{t(b,\tau)}$ is one of right handed
coupling. Since the left handed fermion which belongs to
$SU(2)$ doublet, the couplings of up-side and down-side
in the doublet are the same as each other.

{}From eq.(\ref{3-tc}) the masses of ordinary fermions
are given as,
\bea
  m_t \sim \xi_L^t \xi_R^t \frac{g_{ETC}^2}{M_{ETC}^2} < \bar{U}U >
      \sim \xi_L^t \xi_R^t \frac{g_{ETC}^2}{M_{ETC}^2} 4 \pi F_6^3,
  \\
  m_b \sim \xi_L^t \xi_R^b \frac{g_{ETC}^2}{M_{ETC}^2} < \bar{D}D >
      \sim \xi_L^t \xi_R^b \frac{g_{ETC}^2}{M_{ETC}^2} 4 \pi F_6^3,
  \\
  m_\tau \sim \xi_L^\tau \xi_R^\tau \frac{g_{ETC}^2}{M_{ETC}^2} < \bar{E}E >
      \sim \xi_L^\tau \xi_R^\tau \frac{g_{ETC}^2}{M_{ETC}^2} 4 \pi F_2^3,
\eea
where $M_{ETC}$ is the mass of the ETC gauge boson and $<
\bar{Q}Q >$ is the condensation of technifermions.
$F_{6}$ is the decay constants of technipion in
techniquark sector and $F_2$ is that in technilepton sector. Here
we used the relation of naive dimensional analysis $<
\bar{Q}Q > \sim 4 \pi F_Q^3 $
\cite{georgi}.

%%%%%%%%%%%%%%%%%%%%%%%%%%%%%%%%%%%%%%%%%%%%%%%%%%%%%%%%%%%%%
%\section{Vertex correction}
%%%%%%%%%%%%%%%%%%%%%%%%%%%%%%%%%%%%%%%%%%%%%%%%%%%%%%%%%%
Now, the vertex correction under consideration is shown in
Fig.1(a).  Because we assume that the ETC gauge boson is
much heavier than the weak gauge boson, we can
shrink the gauge propagator as shown in Fig.2.
The ETC interaction in eq.(\ref{3-tc}) becomes the
following effective four-fermi interaction after Fierz
transformation,
\bea
 {\cal L}_{int} =
 &-& \frac{1}{2} {\xi_L^t}^2 \frac{g_{ETC}^2}{M_{ETC}^2}
      ( \bar{q}_L \gamma^\mu \tau^A q_L )
      ( \bar{Q}_L \gamma_\mu \tau^A Q_L )  \nn \\
 &-& \frac{1}{2} {\xi_L^\tau}^2 \frac{g_{ETC}^2}{M_{ETC}^2}
      ( \bar{l}_L \gamma^\mu \tau^A l_L )
      ( \bar{L}_L \gamma_\mu \tau^A L_L ).
\eea
Then we replace the left handed technifermion current by
chiral current \cite{gio}\cite{civ}\cite{kit}\cite{kit2},
that is the Noether current for
$SU(2)_L$ symmetry in our effective
Lagrangian\footnote{The effective lagrangian
include both the technivectormesons
and the techniaxialvectormesons in the techniquark sector and
the neutral technivectormesons and an exsotic charged
left-handed meson in the technilepton sector. Here, for
simplicity, we assume that the masses of the axialvectormesons and the
left-handed meson are much larger than them of the other
vectormeson in each sector. Then, we can ignore the effects
of their technimesons. Indeed,
we have known that the contributions of the
axialvectormesons for $S$ parameter
are smaller than them of the vectormesons in the
previous work \cite{yosi}.
}
 \cite{yosi}. With the replacement, we obtain,
%The effective lagrangian include both the technivectormesons
%and the techniaxialvectormesons in the techniquark sector and
%the neutral technivectormesons and a exsotic charged
%left-handed meson in the technilepton sector. Here, for
%simplicity, we assume that the masses of the axialvectormesons and the
%left-handed meson are much larger than them of the other
%vectormeson in each sector. Then, we can ignore the effects
%of their technimesons and do not need to consider them. In
%deed, we have known that the contributions of the
%axialvectormesons for $S$ parameter
%are smaller than them of the vectormesons in the
%previous work \cite{yosi}.
%In the technilepton sector, from the isospin breaking, there
%are the mixing between neutral vectormesons, techni-$\rho$
%and techni$\omega$ in both mass terms and kinetic terms of
%the vectormesons. So we must diagonalize the mixing, when we
%compute the effects of the technivectormeson in technilepton
%sector.
\bea
 {\cal L}_{int} =
 - \frac{1}{2} {\xi_L^t}^2 \frac{g_{ETC}^2}{M_{ETC}^2}
    [ \bar{q}_L \gamma^\mu \{
            &-& F_6^2 \frac{\sqrt{3}}{2}
               ( g W_\mu^3 - g^\prime B_\mu) \frac{\tau^3}{ \sqrt{3}}
            - F_6^2 \frac{\sqrt{3}}{2}
               \sum_{a=1}^2 ( g W_\mu^a )
                          \frac{\tau^a}{\sqrt{3}}
                          \nn \\
            &+& \frac{M_{V6}^2}{G_6}
                     [ \rho_{6\mu}^3 - \frac{\sqrt{3}}{2 G_{6}}
               ( g W_\mu^3 + g^\prime B_\mu)]\frac{ \tau^3}{ \sqrt{3}}
                        \nn \\
            &+& \frac{M_{\omega 6}^2}{G_{6 \omega}}
                     [ \omega_{6\mu} - \frac{\sqrt{3}}{2 G_{6\omega}}
               2 Y_{Lq} g^\prime B_\mu ]\frac{ Y_{Lq}}{ \sqrt{3}} \}
                                        q_L ]
                                \nn \\
 - \frac{1}{2} {\xi_L^\tau}^2 \frac{g_{ETC}^2}{M_{ETC}^2}
    [ \bar{l}_L \gamma^\mu \{
            &-& F_2^2 \frac{1}{2}
               ( g W_\mu^3 - g^\prime B_\mu) \tau^3
             - F_L^2 \frac{1}{2}
               \sum_{a=1}^2 ( g W_\mu^a ) \tau^a
                        \nn \\
            &+& \frac{M_{V2}^2}{G_2}
                     [ \rho_{2\mu}^3 - \frac{1}{2 G_{2}}
               ( g W_\mu^3 + g^\prime B_\mu)]\tau^3
                        \nn \\
            &+& \frac{M_{\omega 2}^2}{G_{2 \omega}}
                     [ \omega_{2\mu} - \frac{1}{2 G_{6\omega}}
               2 Y_{Ll} g^\prime B_\mu ]Y_{Ll} \nn \\
            &+& \frac{\beta_V}{2 G_2}
                     [ \omega_{2 \mu} - \frac{1}{2 G_{2\omega}}
              2 Y_{Ll} g^\prime B_\mu ] \tau^3 \nn \\
            &+& \frac{\beta_V}{2G_{2\omega}}
                     [\rho_{2\mu}^3 - \frac{1}{2G_2}
              ( g W_\mu^3 + g^\prime B_\mu) ]Y_{Ll}
\}
                                        l_L ] ,
\label{6}
\eea
where, we followed the same notation as that used in
ref.\cite{yosi}.
$\rho$ and $\omega$ are technivectormesons which are the bound
states  of
technifermion. $M_V$ and $M_{\omega}$ are their masses.
In the technilepton sector, because of the presence of
the isospin breaking terms, there
are the mixings between neutral vectormesons, techni-$\rho$
and techni-$\omega$.
Therefore we must diagonalize the mixing when we
compute the effects of the technivectormeson in technilepton
sector.

With eq.( 6 ), the vertex corrections
are the following,
\bea
  \delta g_L^{Zb\bar{b}} &=& \frac{1}{4} {\xi_L^t}^2
             \frac{g_{ETC}^2}{M_{ETC}^2} F_6^2
             \sqrt{ g^2 + {g^\prime}^2 } + \delta{\bar{g}_L^{Zb\bar{b}}},
 \label{9} \\
  \delta g_L^{Z\tau^+ \tau^-} &=& \frac{1}{4} {\xi_L^\tau}^2
             \frac{g_{ETC}^2}{M_{ETC}^2} F_2^2
             \sqrt{ g^2 + {g^\prime}^2 }
                      + \delta{\bar{g}_L^{Z\tau^+\tau^-}}, \\
  \delta g_L^{W\tau \nu} &=& - \frac{1}{2\sqrt{2}}
             {\xi_L^\tau}^2
             \frac{g_{ETC}^2}{M_{ETC}^2} F_L^2 g
                      ,
\label{11}
\eea
where $\delta{\bar{g}_L^{Zb\bar{b}}}$ and
$\delta{\bar{g}_L^{Z\tau\tau}}$ are the corrections from
the effect from the technivectormesons in each sector as shown in Fig.3.
We assume that technivectormesons in techniquark
sector can be ignored when
their masses are very heavy, $M_{V6}$,$M_{\omega 6}$ $\sim$ 1
TeV,
compared with weak
gauge boson masses. On the other hand, it is expected that
the masses of the technivectormesons in the
technilepton sector are lighter than those in the
techniquark sector in this model \cite{apel}, because the
pion decay constants in the technilepton sector are much
smaller than those in the techniquark sector. Then the
contribution of these light technivector mesons may
be large and can not be ignored. Therefore
we also compute the effects of technivectormesons in technilepton
sector.
Substitutiong for $g_{ETC}^2/M_{ETC}^2$ from eq.(3) in
eqs.(\ref{9})-(\ref{11}), we find
\bea
  \delta g_L^{Zb\bar{b}} &=& \frac{1}{4}
             \frac{\xi_L^t}{\xi_R^t}
             \frac{m_t}{4 \pi F_6}
             \sqrt{ g^2 + {g^\prime}^2 },
             \\
  \delta g_L^{Z\tau^+\tau^-} &=& \frac{1}{4}
             \frac{{\xi_L^\tau}^2 }{\xi_L^t \xi_R^t}
             m_t \frac{F_2^2}{4 \pi F_6^3}
             \sqrt{ g^2 + {g^\prime}^2 }
             + \delta{\bar{g}_L^{Z\tau\tau}}, \\
  \delta g_L^{W \tau \nu} &=& -
              \frac{1}{2\sqrt{2}}
              \frac{{\xi_L^\tau}^2 }{\xi_L^t \xi_R^t}
              m_t \frac{F_L^2}{4 \pi F_6^3} g
                       .
\eea
The correction from the vectormesons is,
\bea
\delta {\bar{g}_{L}^{Z \tau \tau}}
    = \frac{1}{4} {\xi_L^\tau}^2 \frac{g_{ETC}^2}{M_{ETC}^2}
      &\{&  \frac{1}{G_2^2}
             [
               \frac{M_\rho^2}{2} A_\rho^2
                       \frac{p^2}{p^2 - M_\rho^2}
             + \frac{M_{\omega}^2}{2} A_\omega^2
                       \frac{p^2}{p^2 - M_{\omega}^2}]
          \frac{g^2 - {g^\prime}^2 }
                    { \sqrt{g^2 + {g^\prime}^2}} \nn \\
      &+& \frac{1}{G_{2\omega}^2}
            [
             \frac{M_\rho^2}{2} B_\rho^2
                      \frac{p^2}{p^2 - M_\rho^2}
            + \frac{M_\omega^2}{2} B_\omega^2
                      \frac{p^2}{p^2 - M_\omega^2}]
          \frac{{2g^\prime}^2}
                    { \sqrt{g^2 + {g^\prime}^2}}  \\
     &+& \frac{1}{G_2 G_{2\omega}}
           [
            \frac{M_\rho^2}{2} A_\rho B_\rho
                     \frac{p^2}{p^2 - M_\rho^2}
          + \frac{M_\omega^2}{2} A_\omega B_\omega
                     \frac{p^2}{p^2 - M_\omega^2}]
             \sqrt{g^2 + {g^\prime}^2}  \}, \label{20} \nn
\eea
with
\bea
   A_\rho = c_V ( 1 - \alpha_V )^{\frac{1}{2}} - s_V ( 1 +
                         \alpha_V )^{\frac{1}{2}}, \label{21}\\
   B_\rho = - c_V ( 1 - \alpha_V )^{\frac{1}{2}} - s_V ( 1 +
                         \alpha_V )^{\frac{1}{2}}, \\
   A_\omega = c_V ( 1 + \alpha_V )^{\frac{1}{2}} + s_V ( 1 -
                           \alpha_V )^{\frac{1}{2}}, \\
   B_\omega = c_V ( 1 + \alpha_V )^{\frac{1}{2}} - s_V ( 1 -
                         \alpha_V )^{\frac{1}{2}}, \label{24}
\eea
where we followed the notation of ref.\cite{yosi}.
$\alpha_V$ is a parameter which indicates the isospin
breaking (the mixing between techni-$\rho$ and
techni-$\omega$ in their kinetic terms of them), and $c_V$
and
$s_V$ represent $cos\theta_V$, $sin\theta_V$, where $\theta_V$
is the mixing angle to diagonalize the $\rho-\omega$ mixing
terms.

%%%%%%%%%%%%%%%%%%%%%%%%%%%%%%%%%%%%%%%%%%%%%%%%%%%%%%%%%%%%%%%%%
%\section{The condition for decay constant of techni-pion}
%%%%%%%%%%%%%%%%%%%%%%%%%%%%%%%%%%%%%%%%%%%%%%%%%%%%%%%%%%%%%%%%
Now, we impose the constraints for the decay constants of technipion
and consider some conditions which satisfy them.
In the present model, in order to satisfy the constraint of
the oblique correction, the pion decay constant in
the technilepton sector must be much smaller than the
decay constant in techniquark sector. We search the
values of decay constant which satisfy the conditions, and
compute the vertex
corrections for $Zbb, Z\tau\tau$ and $W \tau \nu $.
First, we can obtain the constraints from the relation
between  weak-gauge boson masses and
the decay constants. In a one-family technicolor model
with custodial symmetry the
constraint is $4 F_{\pi}^2 \simeq (250)^2$. On the other
hand, because in the present model the decay constant in the
technilepton sector are different from that in the
techniquark sector,
the constraint is
\beq
  3 F_6^2 + F_2^2 \simeq ( 250 )^2.
\label{cons1}
\eeq
The second constraint is obtained from $T$ parameter
\cite{pes} which
indicates the breaking of custodial symmetry. The condition is obtained
from the constraint of $T$ parameter \cite{pes}.  The upper bound of $T$
parameter is
\bea
T < 0.5 .
\label{cons2}
\eea
$T$ parameter is given by \cite{yosi},
\bea
   \alpha T = \frac{ F_L^2 - F_2^2 }{ 3 F_6^2 + F_2^2 }. \nn
\eea
Combining eq.(\ref{cons1}) with eq.(\ref{cons2}),
we obtain the constraint between $F_L$ and $F_2$,
\bea
   F_L^2 - F_2^2 < 300.
\eea
The last constraint is obtained from the ratios of masses
of
ordinary fermions
$m_\tau : m_b : m_t \sim 1 : 3: 100$ .
{}From mass formulae in eqs.(3) - (5), we obtain
\bea
   \xi_L^\tau \xi_R^\tau F_2^3 : \xi_L^t \xi_R^b F_6^3 : \xi_L^t \xi_R^t F_6^3
   \sim 1 : 3 : 100 .
\eea
To determine the decay constants, we need to make some
assumptions on the coupling constans $\xi$s.
Here we assume that the difference between the
masses of the ordinary quark and the lepton comes from the
differences of the decay constants of technipion in each
sector. There are two cases roughly. One of them
is that the difference of the decay constants is due to the
difference  between the
masses of the up-type quark ($t$) and the lepton ($\tau$).
The other is that the difference is due to the
difference between
the down-type quark ($b$) and the lepton. Correspondingly,
we assume the relations among the couplings $\xi$s, i.e.,
(A) $\xi_L^\tau \xi_R^\tau = \xi_L^t \xi_R^t$ and (B)
$\xi_L^\tau \xi_R^\tau = \xi_L^t \xi_R^b$. For both
cases,
 we can determine the values of the pion decay constants
with the constraints on eq.(18), eq.(20) and eq.(21)
.\\
 ( A ) \ \ $\xi_L^\tau \xi_R^\tau = \xi_L^t \xi_R^t$:
 \  \  \  $F_6=143GeV$, $F_2=31GeV$, $F_L=35GeV$ \\
 ( B ) \ \ $\xi_L^\tau \xi_R^\tau = \xi_L^t \xi_R^b$:
 \  \  \  $F_6=135GeV$, $F_2=90GeV$, $F_L=92GeV$ \\
In both cases, we compute the vertex correction for $Zb\bar{b}$,
$Z\tau\tau$ and $W\tau\nu$ without including the correction
due to
the technivectormesons ($\delta\bar{g}$)
as shown Table 1,
Table 2 and Table 3 respectively. For comparison, as
case( C ), we show the vertex correction for the case when
the decay constants in technilepton and techniquark sector
are degenerate.
Here, we find that the contributions of technipion for
their vertex corrections ( $\delta g - \delta\bar{g}$ )
become large, as the difference
between the decay constants in the techniquark sector and
the technilepton sector is becoming
smaller.

%\section{Other effects for the vertex corrections}

Next, we consider the correction,
$\delta{\bar{g}_L^{Z\tau\tau}}$, which comes from the
technivectormesons in the technilepton sector.
For simplicity, we put $\alpha_V \sim 1$, $c_V \sim 1$ and $s_V
\sim 0$ in the factors in eqs.(\ref{21})-(\ref{24}),
and substitute for
$\frac{g_{ETC}^2}{M_{ETC}^2}$ from eq.(2) in eq.(13).
Then eq.(13) becomes,
 \bea
\delta {\bar{g}_{L}^{Z \tau \tau}}
    = \frac{1}{4} \frac{{\xi_L^\tau}^2}{\xi_L^t \xi_R^t}
          \frac{m_t}{4 \pi F_6^3}
      &\{&  \frac{1}{G_2^2}
             [
              M_{\omega}^2
                       \frac{p^2}{p^2 - M_{\omega}^2}]
          \frac{g^2 - {g^\prime}^2 }
                    { \sqrt{g^2 + {g^\prime}^2}} \nn \\
      &+& \frac{1}{G_{2\omega}^2}
            [
            M_\omega^2
                      \frac{p^2}{p^2 - M_\omega^2}]
          \frac{{2g^\prime}^2}
                    { \sqrt{g^2 + {g^\prime}^2}} \nn \\
     &+& \frac{1}{G_2 G_{2\omega}}
           [
           M_\omega^2
                     \frac{p^2}{p^2 - M_\omega^2}]
             \sqrt{g^2 + {g^\prime}^2} \}
\eea
Here at the scale of $p^2 \simeq M_Z^2$, we find
that the
contribution becomes large in the following cases.
(1) The technivectormeson's mass is
close to the gauge-boson's mass. (2) The couplings $G_2$ and $G_{2\omega}$
are becoming smaller. Because the smaller values for
$G_2$ and $G_{2\omega}$ are favored to
satisfy the constraint of the oblique correction $S$
\cite{yosi}, the contribution from the technivectormesons will
also be large. In Fig.4, we present the behavior of
$\delta{{g}_L^{Z\tau\tau}}$ including the contribution of the
technivectormesons $\delta{\bar{g}_L^{Z\tau\tau}}$ as a function of
$M_{\omega}$ for several sets of values of $G_2$ and
$G_{2\omega}$. In the
previous work \cite{yosi}, we find that the value of $G_6^2$
is 31.5 when
the custodial symmetry for a doublet is exact. The upper
bound of $G_{2\omega}$ \cite{yosi} which makes $S$ to be negative is,
\bea
    G_{2\omega} < \frac{G_6}{\sqrt{3}} \sim \frac{5.61}{\sqrt{3}}
\sim 3.24.
\eea
Therefore we plot the graph in the following three cases in Fig.4. \\
(1) \ \ \ $G_2, G_{2\omega} = 5.61$ \\
(2) \ \ \ $G_2, G_{2\omega} = 3.24$ ( $S = 0$ ) \\
(3) \ \ \ $G_2, G_{2\omega} = 2$ ( $S = -2$ ) \\
The case(1) is one with positive $S$ like the traditional
technicolor model with custodial symmetry. The
case(2) is one with $S = 0$, and the case(3) is an extreme
case with $S = -2$.  Here we obtain the suppression on the vertex
correction for $Z \tau \tau$ due to the vector meson when
$S$ is negative (Fig.4). However, the correction for $W\tau\nu$ does
not depend on the effects of the technivectormesons, because
there are not such charged technivectormesons in the present
model.
We find that the difference between the vertex corrections of
$Z\tau\tau$ and $W\tau\nu$ in terms of the contribution from
the technivectormesons
appear. In other words, the difference will be the evidence
of the isospin breaking in technilepton sector.

%%%%%%%%%%%%%%%%%%%%%%%%%%%%%%%%%%%%%%%%%%%%%%%%%%%%%%%%%%%%%%%%%%%
%     \section{Conclusion}
%%%%%%%%%%%%%%%%%%%%%%%%%%%%%%%%%%%%%%%%%%%%%%%%%%%%%%%%%%%%%%%%
In this letter we have described the vertex correction of $Zb\bar{b}$,
$Z\tau\tau$ and $W\nu\tau$ in the one family extended technicolor model
without exact custodial symmetry.
The values of the corrections can not be determined
precisely, since the corrections include a few
unknown parameters $\xi$s. The corrections which are obtained in this
letter is much larger than those in
one doublet model. If we suitably choose each unknown
parameter $\xi$, we will be able to obtain
the vertex corrections which satisfy the constraint from the experiment.
When $\xi = 1$,
the vertex corrections are so large that this model is ruled
out. Then, in order to reduce the values in
this case, we may have to consider
the other ETC model or walking technicolor.
However, we find that if the the difference between the
vertex corrections for $Z\tau\tau$ and $W\tau\nu$ is
measured in experiment, it is
the evidence of the isospin breaking in the
technilepton sector. It comes from not only the difference
between the decay constants but a large contribution to
$Z\tau\tau$ vertex due to the technivectormesons.
The contributions of the vectormesons for
$Z\tau\tau$ reduce the value which takes account of only
technipion contribution (the first term of the eq.(11)).
The vertex correction for the
$Z\tau\tau$ can be negative due to this effect. However, the contribution of
the technivectormesons are absent in the vertex
correction for the $W\tau\nu$. Hence, the difference
between the corrections
for the $Z\tau\tau$ and the $W\tau\nu$ appear.
We expect that in the near future the precession
measurements (in  LEP, JLC etc.) of the
vertex corrections of $W\tau\nu$
will determine whether the isospin of
technilepton sector breaks or not.

{\Large{\bf Acknowledgement}}\\
We would like to thank T.Morozumi and L.T.Handoko for
discussion and reading the manuscript.

\newpage

\begin{center}
{\Large{\bf Table Captions}}
\end{center}
\begin{itemize}
\item {\bf Table 1 }: The value of the vertex correction of
$Zb\bar{b}$ and
an amount of shifting the $Zb\bar{b}$ width from the
standard model
in
a one-family technicolor model without exact custodial
symmetry for
each cases.

\item {\bf Table 2 }: The value of the vertex correction of
$Z\tau\tau$ and
an amount of shifting the $Z\tau\tau$ width from the
standard model except for the contribution from the
technivectormesons
in
a one-family technicolor model without exact custodial
symmetry for
each cases.

\item {\bf Table 3 }: The value of the vertex correction of
$W\tau\nu$ and
an amount of shifting the $W\tau\nu$ width from the
standard model
in
a one-family technicolor model without exact custodial
symmetry for
each cases.

\end{itemize}

\begin{center}
{\Large{\bf Figure Captions}}
\end{center}
\begin{itemize}
\item {\bf Fig. 1(a) }: The Feynman daigram for the
contribution to the vertex correction according to side way
ETC gauge interaction.

\item {\bf Fig. 1(b) }: The Feynman diagram for the
contribution to the vertex correction according to diagonal
ETC gauge interaction.

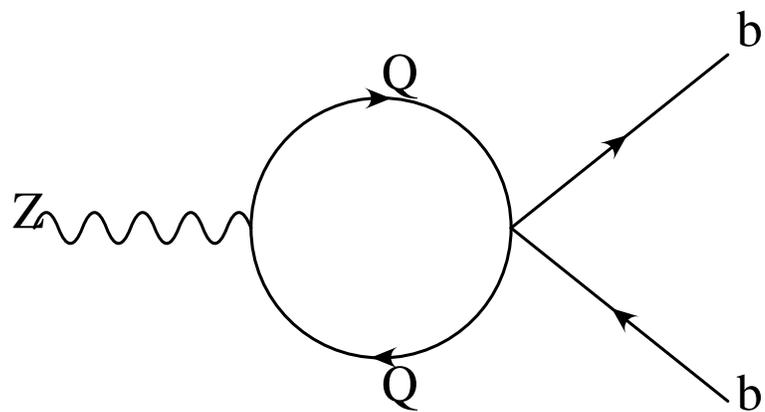
\item {\bf Fig. 2 }: The Feynman diagram in which the ETC
gauge boson propagators are shrunk in Fig.1(a) and Fig.1(b).

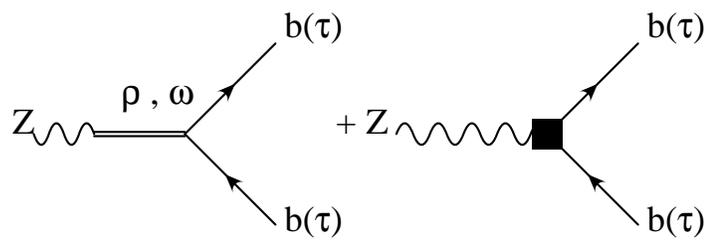
\item {\bf Fig. 3 }: The Feynman diagram to
compute the vertex correction by effective lagrangian
approach. The first shows the effects of
thechnivectormesons. The second shows the effects of the
thecnipion.

\item {\bf Fig. 4 (A) (B) }: In (A) ( $F_6,F_2$ ) =
( 143GeV,31GeV ) and (B) ( $F_6,F_2$ ) = ( 135GeV,90GeV ),
plotting the
$\delta{g_L^{Z\tau\tau}}$ as a function of $M_{\omega}$ for
the each casees, (1) $G_{2\omega} = 5.61$ with a dashline,
(2) $G_{2\omega} = 3.24 \ (S=0)$ with a thickline and (3)
$G_{2\omega} = 2 \ (S=-2)$ with a thinline.

\end{itemize}

\newpage

{\Large
\[ \begin{array}{|c|c|c|c|c|}  \hline
  & (F_6, F_2) & \delta g_L^{Zb\bar{b}}
  & \frac{\delta \Gamma}{\Gamma} \\ \hline
  ( A ) & ( 143GeV, 31GeV )
   & 0.0181 (\frac{m_t}{175})\frac{\xi_L^t}{\xi_R^t}
   & - 11 \% (\frac{m_t}{175})\frac{\xi_L^t}{\xi_R^t}
   \\ \hline
  ( B ) & ( 135GeV, 90GeV )
   & 0.0192 (\frac{m_t}{175})\frac{\xi_L^t}{\xi_R^t}
   & - 11 \% (\frac{m_t}{175})\frac{\xi_L^t}{\xi_R^t}
   \\ \hline
  ( C ) & ( 125GeV, 125GeV )
   & 0.0207 (\frac{m_t}{175})\frac{\xi_L^t}{\xi_R^t}
   & - 13 \% (\frac{m_t}{175})\frac{\xi_L^t}{\xi_R^t}
   \\ \hline
 \end{array}
\]}

\begin{center}
{\bf Table 1}
\end{center}

{\Large
\[ \begin{array}{|c|c|c|c|c|}  \hline
  & (F_6, F_2) & \delta g_L^{Z\tau^+ \tau^-} -
     \delta{\bar{g}_L^{Z\tau\tau}}
  & \frac{\delta \Gamma}{\Gamma} \\ \hline
  ( A ) & ( 143GeV, 31GeV )
   & 0.0009 (\frac{m_t}{175})\frac{{\xi_L^\tau}^2}{\xi_L^t\xi_R^t}
   & - 0.5 \% (\frac{m_t}{175})\frac{{\xi_L^\tau}^2}{\xi_L^t\xi_R^t}
   \\ \hline
  ( B ) & ( 135GeV, 90GeV )
   & 0.0085 (\frac{m_t}{175})\frac{{\xi_L^\tau}^2}{\xi_L^t\xi_R^t}
   & - 4.9 \% (\frac{m_t}{175})\frac{{\xi_L^\tau}^2}{\xi_L^t\xi_R^t}
   \\ \hline
  ( C ) & ( 125GeV, 125GeV )
   & 0.0207 (\frac{m_t}{175})\frac{{\xi_L^\tau}^2}{\xi_L^t\xi_R^t}
   & - 12 \% (\frac{m_t}{175})\frac{{\xi_L^\tau}^2}{\xi_L^t\xi_R^t}
   \\ \hline
 \end{array}
\]}

\begin{center}
{\bf Table 2}
\end{center}

{\Large
\[ \begin{array}{|c|c|c|c|}  \hline
  & (F_6, F_L) & \delta g_L^{W\tau \nu_\tau}
  & \frac{\delta \Gamma}{\Gamma} \\ \hline
  ( A ) & ( 143GeV, 35GeV )
   & -0.0015 (\frac{m_t}{175})\frac{{\xi_L^\tau}^2}{\xi_L^t\xi_R^t}
   & - 0.5 \% (\frac{m_t}{175})\frac{{\xi_L^\tau}^2}{\xi_L^t\xi_R^t}
   \\ \hline
  ( B ) & ( 135GeV, 92GeV )
   & - 0.0126 (\frac{m_t}{175})\frac{{\xi_L^\tau}^2}{\xi_L^t\xi_R^t}
   & - 3.8 \% (\frac{m_t}{175})\frac{{\xi_L^\tau}^2}{\xi_L^t\xi_R^t}
   \\ \hline
  ( C ) & ( 125GeV, 125GeV )
   & - 0.0292 (\frac{m_t}{175})\frac{{\xi_L^\tau}^2}{\xi_L^t\xi_R^t}
   & - 9 \% (\frac{m_t}{175})\frac{{\xi_L^\tau}^2}{\xi_L^t\xi_R^t}
   \\ \hline
 \end{array}
\]}

\begin{center}
{\bf Table 3}
\end{center}

\newpage

% eepic Version 1.1b < Febrary 7, 1988 >
%	Written by Conrad Kwok
%
% Internet : kwok@iris.ucdavis.edu
% csnet    : kwok@ucd.csnet
% csnet    : kwok%iris.ucdavis.edu@csnet.relay
% UUCP	   : ...!ucbvax!ucdavis!iris!kwok
%
% The macros are in public domain.
% You may distribute or modify it in any ways you like.
% Please report any bugs, enhancements, comments, suggestions, etc.
%
% This style file modify some of the commands in epic[1] and LaTeX[2] such
% that \special commands will be generated in drawing lines if approriate.
% The \special commands generated is the subset of the \specials used
% by tpic[3].
%
% [1] epic is written by Sunil Podar. Please read epic.sty for the
%     copyright notice.
% [2] LaTeX is written by Leslie Lamport. Please read the book LaTeX
% [3] tpic is modified from pic by Tim Morgan
%
%% This file contains extensions of the following epic commands:
%%	\dottedline	\dashline	\drawline
%%	\jput
%%
%% It also contains extensions of the following LaTeX commands:
%%	\circle		\line		\oval
%%
%% New commands include:
%%	\Thicklines	\arc		\ellipse
%%	\path		\spline		\allinethickness
%%
%% New commands for eepic 1.1
%%	\blacken	\whiten		\shade
%%	\texture	\filltype{type} type=black|white|shade
%%
%% For documentation, please see the accompanying manual
%%
%% Change Log;
%% o	(October 2, 88)
%% 	Fixed the problem of \line (\@sline). When both x and y are
%% 	large, it produced 'bad character code' error.
%%
%% o    (January 17, 89)
%%	Add commands \blacken, \whiten, \shade, \texture
%%
%% o	(January 17, 89)
%%	Add \filltype{xxxxx}
%%	xxxxx - black, white, shade
%%
\makeatletter
\typeout{%
Extension to Epic and LaTeX. Version 1.1b - Released Febrary 7, 1988}
\newcount\@gphlinewidth
\newcount\@eepictcnt
\newdimen\@tempdimc
\@gphlinewidth\@wholewidth \divide\@gphlinewidth 4736

%% Redefine \thinlines, \thicklines
%% See also latex.tex
\def\thinlines{\let\@linefnt\tenln \let\@circlefnt\tencirc
    \@wholewidth\fontdimen8\tenln \@halfwidth .5\@wholewidth
    \@gphlinewidth\@wholewidth \divide\@gphlinewidth 4736\relax}
\def\thicklines{\let\@linefnt\tenlnw \let\@circlefnt\tencircw
    \@wholewidth\fontdimen8\tenlnw \@halfwidth .5\@wholewidth
    \@gphlinewidth\@wholewidth \divide\@gphlinewidth 4736
    \advance\@gphlinewidth\@ne   % Make the output looks better
    \relax}
%%
%% To indicate whether the dot character is defined in the dotted join
%%     environment or not (\@ifnodotdef)
\newif\if@nodotdef \global\@nodotdeftrue
%%
%% Redefine \dottedjoin
\def\dottedjoin{\global\@jointhemtrue \global\@joinkind=0\relax
  \bgroup\@ifnextchar[{\global\@nodotdeffalse\@idottedjoin}%
                      {\global\@nodotdeftrue\@idottedjoin[\@empty]}}
%%----------------------------------------------------------------------
%% Redefine \jput
\long\def\jput(#1,#2)#3{\@killglue\raise#2\unitlength\hbox to \z@{\hskip
#1\unitlength #3\hss}%
\if@jointhem \if@firstpoint \gdef\x@one{#1} \gdef\y@one{#2}
\global\@firstpointfalse
 \else\ifcase\@joinkind
    \if@nodotdef
        \@spdottedline{\dotgap@join\unitlength}%
(\x@one\unitlength ,\y@one\unitlength)(#1\unitlength,#2\unitlength)
    \else
	\@dottedline[\dotchar@join]{\dotgap@join\unitlength}%
(\x@one\unitlength,\y@one\unitlength)(#1\unitlength,#2\unitlength)
    \fi
	\or\@dashline[\dashlinestretch]{\dashlen@join\unitlength}[\dotgap@join]%
(\x@one,\y@one)(#1,#2)
	\else\@drawline[\drawlinestretch](\x@one,\y@one)(#1,#2)\fi
    \gdef\x@one{#1}\gdef\y@one{#2}%
 \fi
\fi\ignorespaces}
%%
%% Redefine \dottedline to generate special whenever possible.
\def\dottedline{\@ifnextchar [{\@idottedline}{\@ispdottedline}}
%% Similar to \@idottedline but it generate \special
\def\@ispdottedline#1(#2,#3){\@ifnextchar (%
{\@iispdottedline{#1}(#2,#3)}{\relax}}
\def\@iispdottedline#1(#2,#3)(#4,#5){\@spdottedline{#1\unitlength}%
(#2\unitlength,#3\unitlength)(#4\unitlength,#5\unitlength)%
\@ispdottedline{#1}(#4,#5)}
\def\@spdottedline#1(#2,#3)(#4,#5){%
    \@tempcnta \@gphlinewidth\relax
    \advance\@tempcnta by 2     % solely for better output
    \special{pn \the\@tempcnta}%
    \@tempdimc=#2\relax
    \@tempcnta \@tempdimc\relax \advance\@tempcnta 2368 \divide\@tempcnta 4736
    \@tempdimc=#3\relax
    \@tempcntb -\@tempdimc\relax\advance\@tempcntb -2368 \divide\@tempcntb 4736
    \@paspecial{\the\@tempcnta}{\the\@tempcntb}%
    \@tempdimc=#4\relax
    \@tempcnta \@tempdimc\relax \advance\@tempcnta 2368 \divide\@tempcnta 4736
    \@tempdimc=#5\relax
    \@tempcntb -\@tempdimc\relax\advance\@tempcntb -2368 \divide\@tempcntb 4736
    \@paspecial{\the\@tempcnta}{\the\@tempcntb}%
    \@tempdimc=#1\relax
%%
%% Generate the pen width in terms of inch with 3 decimal digit.
    \@tempcnta \@tempdimc\relax \advance\@tempcnta 2368 \divide\@tempcnta 4736
%% \@tempcnta contain the pen width in terms of thousandth of a inch
%% Then it is converted back to inch. I know the way I use is dumb but
%% I cannot think of any better method
    \@tempcntb \@tempcnta\relax \divide\@tempcntb 1000
    \multiply \@tempcntb 1000 \advance\@tempcnta -\@tempcntb
    \divide\@tempcntb 1000
    \ifnum\@tempcnta < 10
        \special{dt \the\@tempcntb.00\the\@tempcnta}%
    \else\ifnum\@tempcnta < 100
        \special{dt \the\@tempcntb.0\the\@tempcnta}%
    \else
        \special{dt \the\@tempcntb.\the\@tempcnta}%
    \fi\fi
    \ignorespaces
}
\def\@iiidashline[#1]#2[#3](#4,#5)(#6,#7){%
\@dashline[#1]{#2\unitlength}[#3](#4,#5)(#6,#7)%
\@iidashline[#1]{#2}[#3](#6,#7)}
%
%% Redefine \@dashline
\long\def\@dashline[#1]#2[#3](#4,#5)(#6,#7){{%
\x@diff=#6\unitlength \advance\x@diff by -#4\unitlength
\y@diff=#7\unitlength \advance\y@diff by -#5\unitlength
%% correction to get actual width since the dash-length as taken in arguement
%% is the center-to-center of the end-points.
\@tempdima=#2\relax \advance\@tempdima -\@wholewidth
\sqrtandstuff{\x@diff}{\y@diff}{\@tempdima}%
\ifnum\num@segments <3 \num@segments=3\fi% min number of dashes I can plot
% is 2, 1 at either end, thus min num@segments is 3 (including 'empty dash').
\@tempdima=\x@diff \@tempdimb=\y@diff
\divide\@tempdimb by\num@segments
\divide\@tempdima by\num@segments
%% ugly if-then-else. If optional dotgap specified, then use it otherwise
%% make a solid dash.
{\ifx#3\@empty \relax
    \ifdim\@tempdima < 0pt \x@diff=-\@tempdima\else\x@diff=\@tempdima\fi
    \ifdim\@tempdimb < 0pt \y@diff=-\@tempdimb\else\y@diff=\@tempdimb\fi
    \global\setbox\@dotbox\hbox{%
                \@absspdrawline(0pt,0pt)(\@tempdima,\@tempdimb)}%
    \else\global\setbox\@dotbox\hbox{%
        \@spdottedline{#3\unitlength}(0pt,0pt)(\@tempdima,\@tempdimb)}%
    \fi}%
\advance\x@diff by -\@tempdima % both have same sign
\advance\y@diff by -\@tempdimb
%
%%here we correct the number of dashes to be put by reducing them
%%appropriately. (num@segments*\@wholewidth) is in some way the slack we
%%have,and division by dash-length gives the reduction. reduction =
%%(2*num@segments*\@wholewidth)/dash-length
%% (num@segments includes empty ones)
\@tempdima=\num@segments\@wholewidth \@tempdima=2\@tempdima
\@tempcnta\@tempdima\relax \@tempdima=#2\relax \@tempdimb=0.5\@tempdima
\@tempcntb\@tempdimb\relax \advance\@tempcnta by \@tempcntb % round-off error
\divide\@tempcnta by\@tempdima \advance\num@segments by -\@tempcnta
\ifnum #1=0 \relax\else\ifnum #1 < -100
  \typeout{***dashline: reduction > -100 percent implies blankness!***}
\else\num@segmentsi=#1 \advance\num@segmentsi by 100
     \multiply\num@segments by\num@segmentsi \divide\num@segments by 100
\fi\fi
\divide\num@segments by 2 % earlier num@segments included 'empty dashes' too.
\ifnum\num@segments >0 % if =0 then don't divide => \x@diff & \y@diff
 \divide\x@diff by\num@segments%   remain same.
 \divide\y@diff by\num@segments
 \advance\num@segments by\@ne %for the last segment for which I subtracted
	 		     %\@tempdima & \@tempdimb from \x@diff & \y@diff
 \else\num@segments=2 % one at each end.
\fi
%%\typeout{num@segments finally = \the\num@segments}
%% equiv to \multiput(#4,#5)(\x@diff,\y@diff){\num@segments}{\copy\@dotbox}
%% with arguements in absolute dimensions.
\@xdim=#4\unitlength \@ydim=#5\unitlength
\@killglue
\loop \ifnum\num@segments > 0
\unskip\raise\@ydim\hbox to\z@{\hskip\@xdim \copy\@dotbox\hss}%
\advance\num@segments \m@ne\advance\@xdim\x@diff\advance\@ydim\y@diff%
\repeat}%
\ignorespaces}
%% redefine \@drawline
%
\def\@drawline[#1](#2,#3)(#4,#5){{%
\@drawitfalse\@horvlinefalse
\ifnum#1 <0 \relax\else\@horvlinetrue\fi
\if@horvline
 \@spdrawline(#2,#3)(#4,#5)
\else\@drawittrue\fi
%%-------------------------------
\if@drawit
\ifnum #1=0 \relax \else\ifnum #1 < -99
  \typeout{***drawline: reduction <= -100 percent implies blankness!***}%
\else\num@segmentsi=#1 \advance\num@segmentsi by 50
     \multiply\num@segmentsi 2
\fi\fi
\@dashline[\num@segmentsi]{10pt}[\@empty](#2,#3)(#4,#5)
\fi}\ignorespaces}% for \if@drawit
\def\@spdrawline(#1,#2)(#3,#4){%
   \@absspdrawline(#1\unitlength,#2\unitlength)(#3\unitlength,#4\unitlength)
   \ignorespaces
}
\def\@absspdrawline(#1,#2)(#3,#4){%
    \special{pn \the\@gphlinewidth}%
    \@tempdimc=#1\relax
    \@tempcnta \@tempdimc\relax \advance\@tempcnta 2368 \divide\@tempcnta 4736
    \@tempdimc=#2\relax
    \@tempcntb -\@tempdimc\relax \advance\@tempcntb -2368 \divide\@tempcntb
4736
    \@paspecial{\the\@tempcnta}{\the\@tempcntb}%
    \@tempdimc=#3\relax
    \@tempcnta\@tempdimc\relax \advance\@tempcnta 2368 \divide\@tempcnta 4736
    \@tempdimc=#4\relax
    \@tempcntb -\@tempdimc\relax \advance\@tempcntb -2368 \divide\@tempcntb
4736
    \@paspecial{\the\@tempcnta}{\the\@tempcntb}%
    \special{fp}%
    \ignorespaces
}
%%----------------------------------------------------------------------
\def\@paspecial#1#2{%
    \special{pa #1 #2}%
}
%%
%% Functions below modify the LaTeX commands and some additional commands
%% are not supported by LaTeX nor epic such as ThickLines and arc.
%%
%% The macros are for use with LaTeX picture environment
%% By including the macro file, you can draw
%%   1) \line in any slope
%%   2) \circle with any radius
%%   3) \ellipse with any x-axis and y-axis length
%%   4) Circular \arc by giving starting and ending angle (in radius)
%%
%% Furthermore you may draw lines in any thickness.
%%
%%
%% Redefine \thinlines, \thicklines and define \Thickline
%% See also latex.tex
\def\Thicklines{\let\@linefnt\tenlnw \let\@circlefnt\tencircw
    \@wholewidth\fontdimen8\tenlnw \@wholewidth 1.5\@wholewidth
    \@halfwidth .5\@wholewidth
    \@gphlinewidth\@wholewidth \divide\@gphlinewidth 4736\relax}
%%
%% Generate the \special command for drawing arc
\def\@circlespecial#1#2#3#4{%
	      \special{pn \the\@gphlinewidth}%
	      \special{ar 0 0 #1 #2 #3 #4}
}
%%
%% Command for drawing an arc. Elliptical arc command can be generated
%% but all iptex program I saw so far does not support that.
\def\@arc#1#2#3#4{%
%% convert TeX dimension to length in terms thousandth of an inch
	\@tempdima #1\unitlength
	\@tempdimb #2\unitlength
        \@tempcnta\@tempdima \advance\@tempcnta 4736 \divide\@tempcnta 9473
	\@tempcntb\@tempdimb \advance\@tempcntb 4736 \divide\@tempcntb 9473
	\setbox\@tempboxa\hbox{%
	    \@circlespecial{\the\@tempcnta}{\the\@tempcntb}{#3}{#4}}%
        \wd\@tempboxa\z@ \box\@tempboxa}
%%
%% Command for drawing Circle and Ellipse in terms of \@arc
%% replace original \circle
\def\circle{%
    \@ifstar{\copy\@filltype\@circle}{\@circle}}
\def\@circle#1{\@arc{#1}{#1}{0}{6.2832}}
\def\ellipse{%
    \@ifstar{\copy\@filltype\@ellipse}{\@ellipse}}
\def\@ellipse#1#2{{\@arc{#1}{#2}{0}{6.2832}}}
\def\arc#1#2#3{\@arc{#1}{#1}{#2}{#3}}
%%
%% Generate \special commands for drawing line
\def\@linespecial#1#2{%
	      \special{pn \the\@gphlinewidth}%
	      \special{pa 0 0}%
	      \special{pa #1 #2}%
	      \special{fp}%
}
%% Replace original \line. Only change is to call \@ssline instead
%% of \@sline. For description, see latex.tex
\def\line(#1,#2)#3{\@xarg #1\relax \@yarg #2\relax
\@linelen=#3\unitlength
\ifnum\@xarg =0 \@vline
  \else \ifnum\@yarg =0 \@hline \else \@ssline\fi
\fi}
\def\@ssline{%
	\ifnum\@xarg< 0
	  \@negargtrue \@xarg -\@xarg \@tempdima -\@linelen
	\else
	  \@negargfalse \@tempdima\@linelen
	\fi
%% truncation is used in arithmetic
	\@tempcnta\@linelen \divide\@tempcnta 4736
        \@yyarg -\@yarg \multiply\@yyarg \@tempcnta \divide\@yyarg\@xarg
 	\if@negarg
	    \@tempcnta -\@tempcnta
	\fi
	\setbox\@linechar\hbox{\@linespecial{\the\@tempcnta}{\the\@yyarg}}%
	\wd\@linechar\@tempdima
	\@clnht\@linelen
        \multiply\@clnht\@yarg
        \divide\@clnht\@xarg
	\ifnum\@yarg< 0
	  \@clnht -\@clnht
	  \ht\@linechar\z@ \dp\@linechar\@clnht
	\else
	  \ht\@linechar\@clnht \dp\@linechar\z@
	\fi
	\box\@linechar
}
%%
%% Replace original \@sline
\def\@sline{%
	\@ssline
%% Codes below (till end of the command) is only necessary
%% when used with \@svector
	\if@negarg
	  \@yyarg -\@yarg
	\else
	  \@yyarg \@yarg
	\fi
	\setbox\@linechar\hbox{\@linefnt\@getlinechar(\@xarg,\@yyarg)}%
	\ifnum\@yarg> 0
	  \let\@upordown\raise
	  \advance\@clnht -\ht\@linechar
	\else
	  \let\@upordown\lower
	\fi
	\if@negarg \kern\wd\@linechar \fi
}
\def\spline(#1,#2){%
    \special{pn \the\@gphlinewidth}%
    \@spline(#1,#2)%
}
\def\@spline(#1,#2){%
    \@tempdima #1\unitlength
    \@tempdimb #2\unitlength
    \@tempcnta \@tempdima \advance\@tempcnta 2368 \divide\@tempcnta 4736
    \@tempcntb -\@tempdimb \advance\@tempcntb -2368 \divide\@tempcntb 4736
    \@paspecial{\the\@tempcnta}{\the\@tempcntb}%
    \@ifnextchar ({\@spline}{\special{sp}}%
}
\def\path(#1,#2){%
    \special{pn \the\@gphlinewidth}%
    \@path(#1,#2)%
}
\def\@path(#1,#2){%
    \@tempdima #1\unitlength
    \@tempdimb #2\unitlength
    \@tempcnta \@tempdima \advance\@tempcnta 2368 \divide\@tempcnta 4736
    \@tempcntb -\@tempdimb \advance\@tempcntb -2368 \divide\@tempcntb 4736
    \@paspecial{\the\@tempcnta}{\the\@tempcntb}%
    \@ifnextchar ({\@path}{\special{fp}}%
}

%%%%%%%%%% Beginning of \oval %%%%%%%%%%%%%%%
\newdimen\maxovaldiam \maxovaldiam 40pt\relax

\def\@oval(#1,#2)[#3]{\begingroup\boxmaxdepth \maxdimen
  \@ovttrue \@ovbtrue \@ovltrue \@ovrtrue
  \@tfor\@tempa :=#3\do{\csname @ov\@tempa false\endcsname}\@ovxx
  #1\unitlength \@ovyy #2\unitlength
  \@tempdimb \ifdim \@ovyy >\@ovxx \@ovxx\else \@ovyy \fi
  \@ovro \ifdim \@tempdimb>\maxovaldiam \maxovaldiam\else\@tempdimb\fi\relax
  \divide \@ovro \tw@
  \@ovdx\@ovxx \divide\@ovdx \tw@
  \@ovdy\@ovyy \divide\@ovdy \tw@
  \setbox\@tempboxa \hbox{%
  \if@ovr \@ovverta\fi
  \if@ovl \kern \@ovxx \@ovvertb\kern -\@ovxx \fi
  \if@ovt \@ovhorz \kern -\@ovxx \fi
  \if@ovb \raise \@ovyy \@ovhorz \fi}
  \ht\@tempboxa\z@ \dp\@tempboxa\z@
  \@put{-\@ovdx}{-\@ovdy}{\box\@tempboxa}%
  \endgroup}

\def\@qcirc#1#2#3#4{%
    \special{pn \the\@gphlinewidth}%
    \@eepictcnt \@gphlinewidth \divide\@eepictcnt 2
    \@tempcnta #1
      \advance\@tempcnta 2368 \divide\@tempcnta 4736
      \advance\@tempcnta\@eepictcnt
    \@tempcntb #2 \divide\@tempcntb 4736 \advance\@tempcntb 2
    \hbox{%

\@qcircspecial{\the\@tempcnta}{-\the\@eepictcnt}{\the\@tempcntb}{#3}{#4}}%
}
\def\@qcircspecial#1#2#3#4#5{\special{ar #1 #2 #3 #3 #4 #5}}

\def\@ovverta{\vbox to \@ovyy{%
    \if@ovb
        \kern \@ovro
        \@qcirc{\@ovro}{\@ovro}{3.1416}{4.7124}\nointerlineskip
    \else
        \kern \@ovdy
    \fi
    \leaders\vrule width \@wholewidth\vfil \nointerlineskip
    \if@ovt
        \@qcirc{\@ovro}{\@ovro}{1.5708}{3.1416}\nointerlineskip
        \kern \@ovro
    \else
        \kern \@ovdy
    \fi
}\kern -\@wholewidth}

\def\@ovvertb{\vbox to \@ovyy{%
    \if@ovb
        \kern \@ovro
        \@qcirc{-\@ovro}{\@ovro}{4.6124}{6.2832}\nointerlineskip
    \else
        \kern \@ovdy
    \fi
    \leaders\vrule width \@wholewidth\vfil \nointerlineskip
    \if@ovt
        \@qcirc{-\@ovro}{\@ovro}{0}{1.6708}\nointerlineskip
        \kern \@ovro
    \else
        \kern \@ovdy
    \fi
}\kern -\@wholewidth}

\def\@ovhorz{\hbox to \@ovxx{%
    \if@ovr \kern \@ovro\else \kern \@ovdx \fi
    \leaders \hrule height \@wholewidth \hfil
    \if@ovl \kern \@ovro\else \kern \@ovdx \fi
    }}
%%%%%%%%% End of \oval %%%%%%%%%%%%%%

\def\allinethickness#1{\let\@linefnt\tenlnw \let\@circlefnt\tencircw
    \@wholewidth #1 \@halfwidth .5\@wholewidth
    \@gphlinewidth\@wholewidth \divide\@gphlinewidth 4736\relax}

\def\blacken{\special{bk}}
\def\whiten{\special{wh}}
\def\shade{\special{sh}}
\def\texture#1{\special{tx #1}\ignorespaces}
\newbox\@filltype
\setbox\@filltype\hbox{\special{bk}}
\def\filltype#1{\@nameuse{ft@#1}}
\def\ft@black{\setbox\@filltype\hbox{\special{bk}}}
\def\ft@white{\setbox\@filltype\hbox{\special{wh}}}
\def\ft@shade{\setbox\@filltype\hbox{\special{sh}}}
\makeatother

\begin{figure}
% GNUPLOT: LaTeX picture using EEPIC macros
\setlength{\unitlength}{0.240900pt}
\begin{picture}(1500,900)(0,0)
\tenrm
\thicklines \path(220,177)(240,177)
\thicklines \path(1436,177)(1416,177)
\thicklines \path(220,304)(240,304)
\thicklines \path(1436,304)(1416,304)
\put(198,304){\makebox(0,0)[r]{-0.008}}
\thicklines \path(220,431)(240,431)
\thicklines \path(1436,431)(1416,431)
\thicklines \path(220,559)(240,559)
\thicklines \path(1436,559)(1416,559)
\put(198,559){\makebox(0,0)[r]{-0.004}}
\thicklines \path(220,686)(240,686)
\thicklines \path(1436,686)(1416,686)
\thicklines \path(220,813)(240,813)
\thicklines \path(1436,813)(1416,813)
\put(198,813){\makebox(0,0)[r]{0}}
\thicklines \path(220,113)(220,133)
\thicklines \path(220,877)(220,857)
\put(220,68){\makebox(0,0){100}}
\thicklines \path(524,113)(524,133)
\thicklines \path(524,877)(524,857)
\thicklines \path(828,113)(828,133)
\thicklines \path(828,877)(828,857)
\put(828,68){\makebox(0,0){200}}
\thicklines \path(1132,113)(1132,133)
\thicklines \path(1132,877)(1132,857)
\thicklines \path(1436,113)(1436,133)
\thicklines \path(1436,877)(1436,857)
\put(1436,68){\makebox(0,0){300}}
\thicklines \path(220,113)(1436,113)(1436,877)(220,877)(220,113)
\put(45,495){\makebox(0,0)[l]{\shortstack{$\delta{{g}^{Z\tau\tau}}$}}}
\put(828,23){\makebox(0,0){$M_{\omega}$ \ \ \  [GeV]}}
\Thicklines
\path(220,279)(220,279)(232,384)(245,454)(257,504)
(269,541)(281,570)(294,592)(306,611)(318,627)(331,640)
(343,651)(355,661)(367,669)(380,677)(392,684)(404,689)
(417,695)(429,700)(441,704)(453,708)(466,711)(478,715)
(490,718)(503,721)(515,723)(527,726)(539,728)(552,730)
(564,732)(576,734)(588,735)(601,737)(613,739)(625,740)
(638,741)(650,743)(662,744)(674,745)(687,746)(699,747)
(711,748)(724,749)(736,750)(748,751)(760,751)(773,752)
(785,753)(797,754)(810,754)(822,755)
\Thicklines \path(822,755)(834,756)(846,756)(859,757)
(871,757)(883,758)(896,758)(908,759)(920,759)(932,760)
(945,760)(957,761)(969,761)(982,762)(994,762)(1006,762)
(1018,763)(1031,763)(1043,763)(1055,764)(1068,764)(1080,764)
(1092,765)(1104,765)(1117,765)(1129,765)(1141,766)(1153,766)
(1166,766)(1178,766)(1190,767)(1203,767)(1215,767)(1227,767)
(1239,768)(1252,768)(1264,768)(1276,768)(1289,768)(1301,769)
(1313,769)(1325,769)(1338,769)(1350,769)(1362,769)(1375,770)
(1387,770)(1399,770)(1411,770)(1424,770)(1436,770)
\thinlines \path(288,113)(294,141)(306,190)(318,231)
(331,265)(343,295)(355,320)(367,343)(380,362)(392,380)
(404,395)(417,409)(429,422)(441,433)(453,443)(466,453)
(478,461)(490,469)(503,477)(515,484)(527,490)(539,496)
(552,501)(564,506)(576,511)(588,516)(601,520)(613,524)
(625,528)(638,531)(650,535)(662,538)(674,541)(687,544)
(699,546)(711,549)(724,551)(736,554)(748,556)(760,558)
(773,560)(785,562)(797,564)(810,566)(822,567)(834,569)
(846,571)(859,572)(871,573)(883,575)(896,576)
\thinlines \path(896,576)(908,578)(920,579)(932,580)
(945,581)(957,582)(969,583)(982,584)(994,585)(1006,586)
(1018,587)(1031,588)(1043,589)(1055,590)(1068,591)
(1080,592)(1092,592)(1104,593)(1117,594)(1129,595)
(1141,595)(1153,596)(1166,597)(1178,597)(1190,598)
(1203,598)(1215,599)(1227,600)(1239,600)(1252,601)
(1264,601)(1276,602)(1289,602)(1301,603)(1313,603)
(1325,604)(1338,604)(1350,605)(1362,605)(1375,605)
(1387,606)(1399,606)(1411,607)(1424,607)(1436,607)
\thinlines \dashline[-10]{25}(220,673)(220,673)
(232,708)(245,732)(257,748)(269,761)(281,770)
(294,778)(306,784)(318,789)(331,794)(343,797)
(355,801)(367,804)(380,806)(392,808)(404,810)
(417,812)(429,814)(441,815)(453,816)(466,818)
(478,819)(490,820)(503,821)(515,821)(527,822)(539,823)
(552,824)(564,824)(576,825)(588,826)(601,826)(613,827)
(625,827)(638,827)(650,828)(662,828)(674,829)(687,829)
(699,829)(711,830)(724,830)(736,830)(748,831)(760,831)
(773,831)(785,831)(797,832)(810,832)(822,832)
\thinlines \dashline[-10]{25}(822,832)(834,832)(846,832)
(859,833)(871,833)(883,833)(896,833)(908,833)(920,834)
(932,834)(945,834)(957,834)(969,834)(982,834)(994,834)
(1006,835)(1018,835)(1031,835)(1043,835)(1055,835)
(1068,835)(1080,835)(1092,835)(1104,835)(1117,835)
(1129,836)(1141,836)(1153,836)(1166,836)(1178,836)
(1190,836)(1203,836)(1215,836)(1227,836)(1239,836)
(1252,836)(1264,836)(1276,836)(1289,837)(1301,837)
(1313,837)(1325,837)(1338,837)(1350,837)(1362,837)
(1375,837)(1387,837)(1399,837)(1411,837)(1424,837)(1436,837)
\end{picture}

\begin{center}
{\bf Fig.4 (a)}
\end{center}

\medskip
\medskip

% GNUPLOT: LaTeX picture using EEPIC macros
\setlength{\unitlength}{0.240900pt}
\begin{picture}(1500,900)(0,0)
\tenrm
\thicklines \path(220,113)(240,113)
\thicklines \path(1436,113)(1416,113)
\put(198,113){\makebox(0,0)[r]{-0.002}}
\thicklines \path(220,189)(240,189)
\thicklines \path(1436,189)(1416,189)
\thicklines \path(220,266)(240,266)
\thicklines \path(1436,266)(1416,266)
\put(198,266){\makebox(0,0)[r]{0}}
\thicklines \path(220,342)(240,342)
\thicklines \path(1436,342)(1416,342)
\thicklines \path(220,419)(240,419)
\thicklines \path(1436,419)(1416,419)
\put(198,419){\makebox(0,0)[r]{0.002}}
\thicklines \path(220,495)(240,495)
\thicklines \path(1436,495)(1416,495)
\thicklines \path(220,571)(240,571)
\thicklines \path(1436,571)(1416,571)
\put(198,571){\makebox(0,0)[r]{0.004}}
\thicklines \path(220,648)(240,648)
\thicklines \path(1436,648)(1416,648)
\thicklines \path(220,724)(240,724)
\thicklines \path(1436,724)(1416,724)
\put(198,724){\makebox(0,0)[r]{0.006}}
\thicklines \path(220,801)(240,801)
\thicklines \path(1436,801)(1416,801)
\thicklines \path(220,877)(240,877)
\thicklines \path(1436,877)(1416,877)
\put(198,877){\makebox(0,0)[r]{0.008}}
\thicklines \path(220,113)(220,133)
\thicklines \path(220,877)(220,857)
\put(220,68){\makebox(0,0){100}}
\thicklines \path(524,113)(524,133)
\thicklines \path(524,877)(524,857)
\thicklines \path(828,113)(828,133)
\thicklines \path(828,877)(828,857)
\put(828,68){\makebox(0,0){200}}
\thicklines \path(1132,113)(1132,133)
\thicklines \path(1132,877)(1132,857)
\thicklines \path(1436,113)(1436,133)
\thicklines \path(1436,877)(1436,857)
\put(1436,68){\makebox(0,0){300}}
\thicklines \path(220,113)(1436,113)(1436,877)(220,877)(220,113)
\put(45,495){\makebox(0,0)[l]{\shortstack{$\delta{{g}^{Z\tau\tau}}$}}}
\put(828,23){\makebox(0,0){$M_{\omega}$ \ \ \  [GeV]}}
\Thicklines
\path(223,113)(232,221)(245,321)(257,392)(269,445)
(281,486)(294,519)(306,545)(318,567)(331,586)(343,602)
(355,616)(367,628)(380,639)(392,648)(404,657)(417,664)
(429,671)(441,677)(453,683)(466,688)(478,693)(490,697)
(503,701)(515,705)(527,708)(539,711)(552,714)(564,717)
(576,720)(588,722)(601,725)(613,727)(625,729)(638,731)
(650,733)(662,734)(674,736)(687,737)(699,739)(711,740)
(724,742)(736,743)(748,744)(760,745)(773,746)(785,747)
(797,748)(810,749)(822,750)(834,751)
\Thicklines \path(834,751)(846,752)(859,753)(871,754)
(883,754)(896,755)(908,756)(920,757)(932,757)(945,758)
(957,758)(969,759)(982,760)(994,760)(1006,761)(1018,761)
(1031,762)(1043,762)(1055,763)(1068,763)(1080,764)
(1092,764)(1104,764)(1117,765)(1129,765)(1141,766)
(1153,766)(1166,766)(1178,767)(1190,767)(1203,767)
(1215,768)(1227,768)(1239,768)(1252,769)(1264,769)
(1276,769)(1289,769)(1301,770)(1313,770)(1325,770)
(1338,770)(1350,771)(1362,771)(1375,771)(1387,771)
(1399,772)(1411,772)(1424,772)(1436,772)
\thinlines \path(349,113)(355,130)(367,162)(380,190)
(392,215)(404,237)(417,257)(429,275)(441,291)(453,306)
(466,319)(478,332)(490,343)(503,353)(515,363)(527,372)
(539,381)(552,388)(564,396)(576,403)(588,409)(601,415)
(613,421)(625,426)(638,431)(650,436)(662,440)(674,445)
(687,449)(699,453)(711,456)(724,460)(736,463)(748,466)
(760,469)(773,472)(785,475)(797,478)(810,480)(822,483)
(834,485)(846,487)(859,489)(871,491)(883,493)(896,495)
(908,497)(920,499)(932,501)(945,502)(957,504)
\thinlines \path(957,504)(969,505)(982,507)(994,508)
(1006,510)(1018,511)(1031,512)(1043,514)(1055,515)
(1068,516)(1080,517)(1092,518)(1104,519)(1117,521)
(1129,522)(1141,523)(1153,523)(1166,524)(1178,525)
(1190,526)(1203,527)(1215,528)(1227,529)(1239,530)
(1252,530)(1264,531)(1276,532)(1289,532)(1301,533)
(1313,534)(1325,534)(1338,535)(1350,536)(1362,536)
(1375,537)(1387,538)(1399,538)(1411,539)(1424,539)
(1436,540)
\thinlines \dashline[-10]{25}(220,634)(220,634)(232,684)
(245,717)(257,741)(269,758)(281,772)(294,783)(306,792)
(318,799)(331,805)(343,811)(355,815)(367,819)(380,823)
(392,826)(404,829)(417,832)(429,834)(441,836)(453,838)
(466,839)(478,841)(490,842)(503,844)(515,845)(527,846)
(539,847)(552,848)(564,849)(576,850)(588,851)(601,852)
(613,852)(625,853)(638,854)(650,854)(662,855)(674,855)
(687,856)(699,856)(711,857)(724,857)(736,858)(748,858)
(760,859)(773,859)(785,859)(797,860)(810,860)(822,860)
\thinlines \dashline[-10]{25}(822,860)(834,861)(846,861)
(859,861)(871,861)(883,862)(896,862)(908,862)(920,862)
(932,863)(945,863)(957,863)(969,863)(982,863)(994,863)
(1006,864)(1018,864)(1031,864)(1043,864)(1055,864)
(1068,864)(1080,865)(1092,865)(1104,865)(1117,865)
(1129,865)(1141,865)(1153,865)(1166,866)(1178,866)
(1190,866)(1203,866)(1215,866)(1227,866)(1239,866)
(1252,866)(1264,866)(1276,866)(1289,867)(1301,867)
(1313,867)(1325,867)(1338,867)(1350,867)(1362,867)
(1375,867)(1387,867)(1399,867)(1411,867)(1424,867)(1436,867)
\end{picture}

\begin{center}
{\bf Fig.4 (b)}
\end{center}

\end{figure}

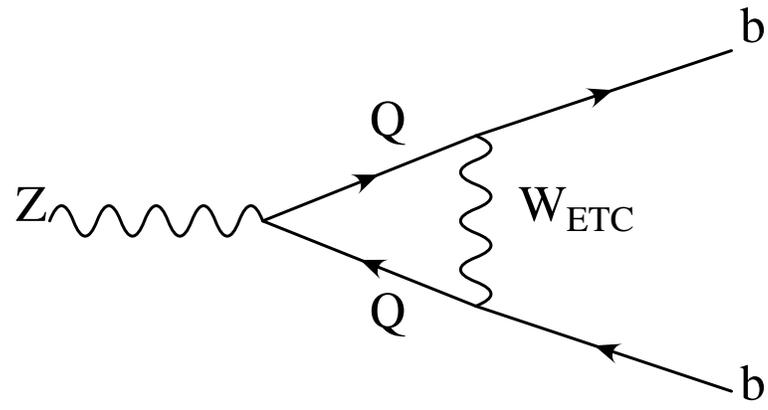
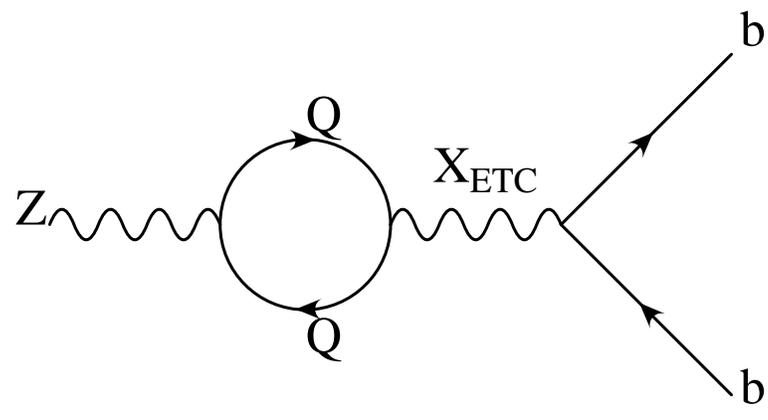
\end{document}